\documentclass[12pt]{iopart}
\usepackage{iopams}

\usepackage{ulem}
\usepackage{times}

\expandafter\let\csname equation*\endcsname\relax
\expandafter\let\csname endequation*\endcsname\relax 

\usepackage[mathscr]{eucal}
\usepackage{amsmath,amsfonts,amssymb,amsthm}
\usepackage{hyperref}
\usepackage{pdfsync}

\newcommand{\ul}[1]{{\underline{#1}}}

\def\sd{S^\dagger}
\def\bsd{\bar S^\dagger}
\def\bs{\bar S}

\def\amb{\fR^{d,2}\backslash\{0\}}

\def\BGST{Barnich:2004cr}
\def\BGadS{Barnich:2006pc}

\def\BGnlp{Barnich:2010sw}
\def\BG-Poincare{Barnich:2009jy}
\def\Fedosov-book{Fedosov:1996fu}

\renewcommand{\tilde}{\widetilde}
\renewcommand{\hat}{\widehat}

\newcommand{\bref}[1]{\textbf{\ref{#1}}}

\newcommand{\gh}[1]{\mathrm{gh}(#1)}

\newcommand{\algf}{\Liealg{f}}
\newcommand{\algh}{\Liealg{h}}
\newcommand{\algg}{\Liealg{g}}

\newcommand{\Liealg}{\mathfrak}       

\newcommand{\dd}{\partial}
\renewcommand{\d}{\partial}

\renewcommand{\geq}{\,{\geqslant}\,}
\renewcommand{\leq}{\,{\leqslant}\,}

\newcommand{\binner}[2]{%
  {\langle}\kern-4.15pt{\langle}#1{,}\,#2{\rangle}\kern-4.15pt{\rangle}}
\newcommand{\commut}[2]{[#1{,}\,#2]}

\newcommand{\half}{\mathchoice{%
    \ffrac{1}{2}}{\frac{1}{2}}{\frac{1}{2}}{\frac{1}{2}}}

\newcommand{\ffrac}[2]{\raisebox{.5pt}%
  {\footnotesize$\displaystyle\frac{#1}{#2}$}\kern1pt}

\newcommand{\dl}[1]{\mathchoice{\ffrac{\dd}{\dd #1}}{\frac{\dd}{\dd
      #1}}{\ffrac{\dd}{\dd #1}}{\ffrac{\dd}{\dd #1}}}

\def\const{\mathop\mathrm{const}\nolimits}
\def\rank{\mathop\mathrm{rank}\nolimits}

\newcommand{\fR}{\mathbb{R}}

\newcommand{\bundle}{\boldsymbol}

\newcommand{\derham}{\boldsymbol{d}}

\newcommand{\manifold}[1]{\mathscr{#1}}
\newcommand{\manX}{\manifold{X}}

\numberwithin{equation}{section} \makeatletter
\@addtoreset{equation}{section}

\begin{document}

\title[Notes on the ambient approach to boundary values of AdS gauge fields]{Notes on the ambient approach to boundary values\\ of AdS gauge fields}

\author{Xavier Bekaert$^a$ and Maxim Grigoriev$^b$}

\address{$^a$~Laboratoire de Math\'ematiques et Physique Th\'eorique\\
    Unit\'e Mixte de Recherche 7350 du CNRS, F\'ed\'eration Denis Poisson\\
    Universit\'e Fran\c cois Rabelais\\
    Parc de Grandmont, 37200 Tours, France}

\address{$^b$~Tamm Theory Department\\
 Lebedev Physics Institute\\ 
 Leninsky prospect 53, 119991 Moscow, Russia}

\begin{abstract}
The ambient space $\fR^{d,2}$ allows to formulate both fields on $AdS_{d+1}$ and conformal fields in $d$ dimensions such that the symmetry algebra $\mathfrak{o}(d,2)$ is realized linearly. We elaborate an ambient approach to the boundary analysis of gauge fields on $AdS_{d+1}$ spacetime.
More technically, we use its parent extension where fields are still defined on $AdS$ or conformal space through arbitrary intrinsic coordinates while the ambient construction works in the target space.
In this way, a manifestly local and $\mathfrak{o}(d,2)$-covariant formulation of the boundary behaviour of massless symmetric tensor gauge fields on $AdS_{d+1}$ spacetime is obtained.
As a byproduct, we identify some useful ambient formulation for Fronsdal fields, conformal currents and
shadow fields along with a concise generating-function formulation of the Fradkin--Tseytlin conformal fields
somewhat similar to the one obtained by Metsaev. We also show how this approach extends to
more general gauge theories and discuss its relation to the unfolded derivation of the boundary dynamics
recently proposed by Vasiliev.

\end{abstract}

\submitto{\JPA}

\section{Introduction}
\label{sec:introduction}

In conformal field theories (CFTs) in spacetime dimension $d>2$, two closely related fields play an important role: the conformal currents and their shadows.
A symmetric tensor field of rank $s$ and conformal weight $\Delta = s+d-2$
which is both divergenceless and traceless is referred to as a ``conformal current'' (with canonical dimension)
while a ``shadow field''\footnote{Through the present paper, we follow the terminology of \cite{Metsaev:2008fs} and reserve the term `shadow field' only for the `shadow' (in the original sense of \cite{Ferrara:1972uq}) of a conformal current.} is the equivalence class of a traceless symmetric tensor field of rank $s$ and (conjugate) conformal weight $\Delta = 2-s$ modulo pure gauge fields (\textit{i.e.} modulo symmetrized traceless gradients).
For spin $s\geqslant 1$, conformal currents are those primary fields that saturate the unitarity bound on the conformal dimension. The shadow fields are those primary fields used in the construction of conformal (higher-spin) gravity theories \cite{Fradkin:1985am,Segal:2002gd} (see \cite{Vasiliev:2009ck} for further generalizations).
They should also be useful in the computation of conformal blocks for currents (see \textit{e.g.} \cite{Costa:2011mg} and references therein).
Moreover, in the generating functional of conformal current correlators the shadow fields play the role of external sources coupling to these primary operators.
Incidentally, in the $AdS/CFT$ correspondence the conformal currents and the shadow fields manifest
themselves in two related ways (see \textit{e.g.} \cite{Balasubramanian:1998sn,Metsaev:1999ui}): the conformal currents appear as boundary values of ``normalizable'' solutions of Fronsdal's equations \cite{Fronsdal:1978vb} for massless symmetric tensor fields on $AdS_{d+1}$, while the shadow fields appear as
boundary values of (so-called) ``non-normalizable'' solutions of the same equations, \textit{i.e.} for different near-boundary behavior. Here, these solutions of Fronsdal's equations will be called respectively ``(non) normalizable Fronsdal fields'' for the sake of brevity. The light-cone and Stuckelberg-like formulations of the aforementioned aspects have been intensively developed by Metsaev over the years \cite{Metsaev:1999ui,Metsaev:2008fs,Metsaev:2009ym}.
The deep relationship between these pair of conformal fields (currents and shadows) and their AdS duals (Fronsdal fields) plays an important role in the conjectured duality between free or critical large-N vector models and higher-spin gravity \cite{Sezgin:2002rt}. This conjecture recently attracted a lot of attention (see \textit{e.g.} \cite{Bekaert:2012ux} for a short review) and provides a strong motivation for the development of various formulations of the above conformal and AdS fields.
Recently, Vasiliev investigated the holographic reduction of his unfolded equations \cite{Vasiliev:1988xc} describing interacting massless fields on $AdS_4$ and obtained nonlinear system for conformal currents and shadow fields in $d=3$ \cite{Vasiliev:2012vf}.

From a group-theoretical perspective, the conformal current and the shadow field modules (\textit{i.e.} the space spanned by these conformal primary fields and all their descendants) and the Fronsdal fields span intertwined $\mathfrak{o}(d,2)$-modules for any given spin. Generically the integral kernel of such intertwiners have direct physical interpretation \cite{Dobrev:1998md}: the two-point correlators define the intertwiners between conformal fields and their shadows while the Witten propagators define the intertwiner from boundary to bulk fields. A subtle point is that, generically, the conformal current and the shadow field are not equivalent as $\mathfrak{o}(d,2)$-modules. Indeed, the conformal current generates a unitary irreducible $\mathfrak{o}(d,2)$-module while, on the contrary, for spin $s\geqslant 1$ the shadow field generates an indecomposable $\mathfrak{o}(d,2)$-module which is reducible (indeed the Weyl-like tensor field built out of a shadow field is a conformal primary) and 
non-unitarizable (since $2-s$ is below the unitarity bound). For spin $s=0$, the situation is somewhat different because, for instance, both modules are irreducible for $d$ odd. Moreover, the scalar shadow field module is unitarizable in dimension $d\leqslant 6$, a phenomenon underlying the holographic degeneracy in the conjecture \cite{Sezgin:2002rt}.
In even dimension $d\geqslant 4$ and for fixed spin, a somewhat confusing point is that the conformal current module appears as a submodule of the corresponding shadow field module. Indeed, the left-hand-side of the Fradkin-Tseytlin equations \cite{Fradkin:1985am} (and their higher-dimensional generalization \cite{Segal:2002gd}) is a descendent of the shadow field and obeys to all the properties of a conformal current \cite{Shaynkman:2004vu}. 
The shadow field module quotiented by the left-hand-side of the Fradkin-Tseytlin equations (and its descendants) will be referred to as ``Fradkin-Tseytlin'' field (module) for short. It is spanned by the inequivalent solutions of Fradkin-Tseytlin equations (of order $d+2s-4$) and generically (for sufficiently high spin $s$ or dimension $d$) it is non-unitary since the equations are higher-derivative. In a sense, a Fradkin-Tseytlin field is an on-shell shadow field.

A celebrated idea, which dates back to Dirac~\cite{Dirac:1935zz}, is to describe AdS and conformal fields in terms of an ambient space $\fR^{d,2}$, often called ``embedding space'' as well, in order to make $O(d,2)$ symmetry manifest in the sense that the group $O(d,2)$ acts linearly on the Cartesian coordinates $X$ for $\fR^{d,2}$. In this approach, the spacetime $AdS_{d+1}$ of curvature radius $R$ is the one-sheeted hyperboloid $X^2=-R^2$ whose conformal
$d$-dimensional boundary is represented as the quotient of the hypercone $X^2=0$ modulo the equivalence relation $X\sim \lambda\, X$ ($\lambda\neq 0$) instead of its usual description as compactified Minkowski spacetime.
In this way, the linear action of $O(d,2)$ on $\fR^{d,2}$
gives the isometry (conformal) group action on anti de Sitter (respectively, compactified Minkowski) spacetime.
This allows to unify AdS and conformal fields as ambient fields defined on $\fR^{d,2}$. The ambient space approach  \textit{\`a la} Fefferman-Graham to conformal geometry and to boundary value problems has been applied to the holographic correspondence for the metric tensor since the early days of the AdS/CFT correspondence (see \textit{e.g.} the review \cite{Alvarez:2002di} and references therein).

The apparent disadvantage of the ambient approach is the lack of transparent
locality in the sense that local (conformal) field theories on
$(\partial)AdS_{d+1}$ spaces are formulated in terms of fields on the ambient
space $\fR^{d,2}$. This can be partially overcome by considering
$\mathfrak{o}(d,2)$-tensor fields defined in terms of the intrinsic geometry of
$(\partial)AdS_{d+1}$ spaces. Such a formulation has been developed
in~\cite{Stelle:1979aj} where the intrinsic geometry is described in terms of an
$\mathfrak{o}(d,2)$-connection and a compensator field. This construction is
known to conformal geometers as the tractor bundle (see \textit{e.g.}
\cite{Gover:2008sw} and references therein). The tractor bundle technique has
been also successfully employed~\cite{Gover:2011rz} in studying boundary
values. 

In this work we develop an ambient space approach to boundary values of
$AdS_{d+1}$ higher-spin gauge fields. Although we explicitly concentrate on
Fronsdal fields, the method is quite general and can be extended to more general
fields on $AdS_{d+1}$ as well as nonlinear gauge theories. Unlike the intrinsic
analysis, in this framework the conformal boundary can be identified as a
submanifold of the ambient space rather than the asymptotic boundary. In
particular, the choice of asymptotic behaviour corresponds to the choice of the
admissible homogeneity degree in the ambient representation. In this way a given
AdS gauge field in general produces two different ambient  (gauge) fields and
hence conformal fields on the boundary. More technically, our analysis is based
on an ambient space construction in the fiber rather than in the base manifold,
in the same spirit as the unfolded description of AdS massless higher-spin
fields in any dimension \cite{Vasiliev:2003ev} (for a review, see \textit{e.g.}
\cite{Bekaert:2005vh}).

In contrast to the standard approach where the boundary value of the Fronsdal field with either shadow-type or current-type asymptotics is off-shell this is not always the case in our framework. More precisely, for even $d$ (and hence odd-dimensional $AdS$-space) the ambient system associated to shadow-type asymptotics simultaneously describe
the conformal field subject to Fradkin--Tseytlin (FT) equations and the conserved conformal current so that both boundary values are encoded into a single conformal system. This is a nonstandard manifestation of the well-known logarithmic term~\cite{Skenderis:2002wp} in the near-boundary expansion.~\footnote{Recall that in the standard approach~(see \textit{e.g.}~\cite{Liu:1998bu,Segal:2002gd,Metsaev:2009ym,BJM} and references therein) FT action is found in the logarithmically-divergent part of the effective action.} Namely, in the unfolded-type framework we are using  there is no room for the logarithmic terms to cancel the anomaly 
(but the formulation can be modified to include such terms) so that for even $d$ the extension of boundary data into the bulk is obstructed and results in the FT equations.  This observation is expected to play an important role at the nonlinear level where both boundary values start to see each other as one can observe analyzing their gauge transformations. 

The plan of our paper is as follows: The ambient approach to AdS fields (massive scalars and Fronsdal fields) 
and to the conformal fields (scalar singletons, conformal currents and shadow fields) is presented in Section \bref{Embedding}. In Section~\bref{sec:amb-app} we explain how the ambient space can be used to study boundary data.
We then introduce in Section \bref{sec:conf-part} our main technical tool: the parent formulation which allows to explicitly relate AdS and conformal fields by lifting the ambient space construction to the target space. The passage from AdS to its boundary roughly amounts to replacing the AdS covariant derivative with the conformal one
and to changing accordingly the compensator gender from time-like to light-like.
The detailed analysis of totally symmetric massless fields is then performed in Section~\bref{sec:boundary}.
In the concluding section~\bref{sec:gen} we show how our framework extends to more general setting and briefly discuss
its relation to the unfolded approach to boundary values.

\section{Ambient approach to AdS and conformal fields}
\label{Embedding}

Let $\fR^{d,2}\backslash\{0\}$ be the pseudo-Euclidean space (with the origin excluded) endowed with the flat metric $\eta_{AB}$ of signature $(--++\ldots+)$. Let $X^A$ ($\,A=+,-,0,1,2,\ldots,d-1\,$) be the standard light-cone coordinates on $\fR^{d,2}$ so that $\eta_{+-}=1=\eta_{-+}$ and $\eta_{ab}=$ diag$(-1,+1,\ldots,+1)$ ($\,a,b=0,1,2,\ldots,d-1\,$).
The $AdS_{d+1}$ spacetime can be seen as the hyperboloid $\eta_{AB}X^AX^B=-R^2$ of radius $R$. In its turn, the $d$-dimensional conformal space $\manX_d$ can be identified with the projective hypercone of light-like rays (see the introduction) and as the conformal boundary of the AdS spacetime\,: $\manX_d\cong \partial ( AdS_{d+1})$.
It can also be seen as the conformal compactification of the Minkowski spacetime $\fR^{d-1,1}$ with Cartesian coordinates $x^a$. Concretely, the Minkowski spacetime $\fR^{d-1,1}$ can be identified with the paraboloid intersection between the null hypercone ($X^2=0$) and a null plane of $\fR^{d,2}\backslash\{0\}$ (say $X^+=1$ and $X^a=x^a$).

\subsection{Ambient representation of the AdS scalar field}

Let $\phi$ denote a scalar field on $AdS_{d+1}$ of mass $m$ satisfying, in intrinsic terms, the Klein-Gordon equation
\begin{equation}
\label{ads-scalar}
 (\nabla^2-m^2)\phi=0\,.
\end{equation}
Equivalently, this can be written in terms of the ambient scalar field $\Phi(X)$
satisfying
\begin{equation}
\label{adscft}
 \left(X \cdot \dl{X}+\Delta\right)\Phi=0\,, \qquad \left(\dl{X}\cdot \dl{X}\right)\Phi=0\,,
\end{equation}
where the homogeneity degree $-\Delta$ in $X$ is related to the mass $m$
through the standard AdS/CFT relation
$m^2=\frac{\Delta(\Delta-d)}{R^2}$, 
so that there are two possible values $\Delta_\pm=\frac{d}{2}\pm\sqrt{(\frac{d}{2})^2+(m R)^2}$ for the
homogeneity degree. As one can see explicitly, the Breitenlohner-Freedman bound  $m^2\geqslant -(\frac{d}{2\,R})^2$
on the mass square is equivalent to the reality of the dual conformal weights $\Delta_\pm$ (and thus of the homogeneity degrees $-\Delta_\pm$).

For $\Phi$ satisfying \eqref{adscft}, one can check that its value on the hyperboloid indeed satisfies~\eqref{ads-scalar} and that (at least locally) any $\phi$ satisfying~\eqref{ads-scalar}
can be lifted to $\Phi$. Indeed, this is just a pseudo-Euclidean
version of the definition of harmonic functions on the sphere $S^{d+1}$ as homogeneous functions on the ambient Euclidean space ${\mathbb R}^{d+2}$.
More precisely, the correspondence between solutions to~\eqref{ads-scalar} and \eqref{adscft} is one-to-one if and only if one restricts to the domain $X^2<0$ of the ambient space. Indeed, the homogeneity constraint $(X \cdot \dl{X}+\Delta)\Phi=0$
defines a unique extension $\Phi$ of $\phi$ to the domain $X^2<0$. This can be easily seen by introducing a radial coordinate $r=\sqrt{-X^2}$ on this domain and considering $\Phi=(\frac{r}{R})^{-\Delta}\phi$.
One can of course try to extend $\Phi$ to a smooth homogeneous
function on the entire ambient space $\fR^{d,2}\backslash \{ 0 \}$ but such an extension is not unique (and not
even guaranteed). This subtlety is important for studying boundary behavior and we discuss it in more details in Section~\bref{sec:amb-app}.

\subsection{Ambient representation of the conformal scalar field and its $\mathfrak{sp}(2)$ algebra of constraints}

The ambient representation of the conformal scalar is based on the following constraints which span an
$\mathfrak{sp}(2)$ algebra 
\begin{equation}
\label{sp2}
 X^2\,,\qquad X \cdot \dl{X}+\frac{d+2}{2}\,, \qquad \dl{X}\cdot \dl{X}
\end{equation}
These operators commute with the operators $L_{AB}=X_{A} \dl{X^{B}}-X_{B} \dl{X^{A}}$ that span 
the conformal algebra $\mathfrak{o}(d,2)$. These two algebras form a Howe dual pair~\cite{Howe} on the space of functions in $X^A$. 

Constraints \eqref{sp2} are extensively
used in the so-called two-time physics \cite{Bars:1998ph} and
in nonlinear higher-spin {gauge theory}~\cite{Vasiliev:2003ev}.
When ``imposed'' on the ambient field $\Phi(X)$ as (strictly speaking, the 2nd and the 3rd constraints are imposed
while the 1st one implements a gauge equivalence {relation})\\
\begin{equation}\label{3cstrs}
\Phi\,\sim\,\Phi\,+\,X^2\,\chi\,, \qquad \left(X\cdot\dl{X}+\frac{d-2}{2}\right)\Phi=0\,,\qquad \left(\dl{X}\cdot \dl{X}\right) \Phi=0\,,
\end{equation}
where $\chi(X)$ is a gauge parameter, 
the {constraints \eqref{sp2}} define \cite{Marnelius:1978fs} the massless scalar field $\phi(x)$ of canonical dimension $\Delta=\frac{d-2}2$, \textit{i.e.} the scalar singleton (see \textit{e.g.} \cite{Bekaert:2009fg} and references therein for the case of generic integer spin).
Indeed, the second condition in \eqref{3cstrs} says that the ambient scalar field $\Phi(X)$ is a homogeneous function of degree $-\Delta=\frac{2-d}2$ so that the evaluation of $\Phi(X)$ on the hypercone $X^2=0$ defines a density $\phi_0(x)$ on $\manX_d$ of conformal weight $\frac{d-2}2$ via the expression $\Phi(X^+,-\frac12 X^+x^2,X^+x^a)=(X^+)^{\frac{2-d}2}\phi_0(x)$ valid in the region $X^+\neq 0$. Conversely, the equivalence relation in \eqref{3cstrs} states that any homogeneous extension of such a given density $\phi_0(x)$ defines a physically equivalent ambient field $\Phi(X)$.
Therefore the scalar singleton can either be seen as a massless scalar field $\phi_0(x)$ living on the conformal
space $\manX_d$ or, equivalently, as a massive scalar field on $AdS_{d+1}$ with specific mass-square $m^2=\frac{4-d^2}{(2R)^2}$ such that the modes corresponding to $\Delta_+=\frac{d+2}2$ boundary behavior form a submodule of solutions that can be quotiented away so that only the modes with $\Delta_-=\frac{d-2}2$ behavior remain (see \textit{e.g.} \cite{Flato:1986uh} for the $d=3$ case). Notice that in the ambient formulation, the latter quotient precisely corresponds to the equivalence relation in \eqref{3cstrs}. So the ambient formulation \eqref{3cstrs} somehow unifies these two celebrated descriptions of the singleton. 

\subsection{Generic implementation of constraints}

For later use, let us define in a precise way what do we mean under ``imposing'' some of the constraints
and gauging away others. Let $\algf$ be an algebra of constraints $T_I$ which are operators acting on a certain linear space (= representation space). For simplicity we assume that all $T_I$ are bosonic and satisfy 
Lie algebra relations $\commut{T_I}{T_J}=f_{IJ}^K T_K$ with $f_{IJ}^K$ the structure constants. Let in addition
$\algf$ be a direct sum (as a linear space, not necessarily as an algebra) of two 
{sub}algebras $\algh\subset \algf$ and $\algg \subset \algf$. In what follows we use the basis $\{T_i, T_\alpha\}$ such that $\{T_i\}$
form a basis in $\algh$ and $\{T_\alpha\}$ in $\algg$. 

This data naturally defines a gauge system for which the constraints $T_i$ give rise to equations of motion while the constraints $T_\alpha$
generate gauge symmetries. An efficient way to explicitly identify equations, gauge symmetries, and constraints for gauge parameters is to employ the BRST technique. Namely, 
to each gauge generator $T_\alpha$ one associates a fermionic ghost variable $b_\alpha$
and introduce ghost degree such that $\gh{b_\alpha}=-1$ and the degree of any other variable vanishes.
This enlarges the representation space by tensoring with the Grassmann algebra generated by $b_\alpha$.
On the extended space, one then introduces the BRST operator
\begin{equation}
 Q=T_\alpha\dl{b_\alpha}-\half  \dl{b_\alpha} \dl{b_\beta}  f^{\gamma}_{\alpha\beta} b_\gamma
\end{equation} 
which is nilpotent and carries ghost degree $1$.

Given such $Q$ one then builds the BRST invariant extensions of the constraints $T_i$\,:
$\hat T_i=T_i+\dl{b_\alpha}  f_{\alpha i}^\beta b_\beta$\,, such that
$\commut{Q}{\hat T_i}=f_{\alpha i}^j \dl{b_\alpha}\hat T_j \,$.
Note that $\commut{\hat T_i}{\hat T_j}=f_{ij}^k \hat T_k$.
The equations of motion, gauge transformation{s}, and constraints for gauge parameters can be then represented as
\begin{equation}
  \hat T_i\Phi=0\,,\qquad \delta_\chi \Phi=Q \chi \,,\qquad \hat T_i \chi=0\,,
\end{equation} 
where $\gh{\Phi}=0$ and $\gh{\chi}=1$ so that $\Phi$ is $b$-independent while  $\chi=b_\alpha \lambda^\alpha$. In components one gets
\begin{equation}
\label{constr4fields-param}
 (T_i+f^\alpha_{\alpha i} )\Phi=0\,, \qquad  
 \delta_\lambda\Phi=T_\alpha \lambda^\alpha-f_{\alpha\beta}^\beta \lambda^\alpha\,,\qquad 
 T_i\lambda^\alpha + f_{\gamma i}^\gamma \lambda^\alpha-f^\alpha_{\gamma i}\lambda^\gamma=0\,.
\end{equation}
Let us stress that according to the first equation the equations of motion in general differ from $T_i \Phi=0$ by a
constant determined by the structure constants of $\algf$. We will see various examples
of this phenomenon.

In the above considerations we used the coordinate representation for the ghost momenta $b_\alpha$. Of course one could have used instead more conventional momenta representation for $b_\alpha$ where $b_\alpha$ and $\dl{b_\alpha}$
are respectively represented as $\dl{c^\alpha}$ and $c^\alpha$ on the space of states depending on ghost coordinates $c^\alpha$ with $\gh{c^\alpha}=1$. If constraints are bosonic this is equivalent and the only difference would be that one would  have to take $\gh{\Phi}=$dim$(\algg)$. If fermionic constraints are present or the algebra is infinite-dimensional, then the equivalence is broken and one is forced to use the representation as above.

Let us also mention that an alternative (and apparently more fundamental) approach is to start with the BRST
operator implementing all the constraints $T_I$, \textit{i.e.} in addition to ghost $b_\alpha$  introduce ghosts $c^i$
with $\gh{c^i}=1$. In this way one arrives at the genuine gauge formulation with unconstrained gauge parameters. 
However, in various applications it often turns out to be useful (see e.g.~\cite{Alkalaev:2008gi,Alkalaev:2009vm}) to employ ``partial'' BRST operator implementing only a subalgebra of the entire constraint algebra and to impose the rest of the constraints by hands.

Using the conformal scalar field as a simple example note that $\algg$ is one-dimensional since it is spanned by the 1st constraint in~\eqref{sp2}. The remaining two constraints in~\eqref{sp2} form subalgebra $\algh$. It is then easy to check that the first two relations of~\eqref{constr4fields-param} indeed give the gauge transformation and 
the constraint~\eqref{3cstrs} including the shift in the ordering constant in the homogeneity constraint.

\subsection{The ambient symmetric tensor fields and their $\mathfrak{sp}(4)$ algebra of constraints}

Consider symmetric tensor fields $\Phi_{A_1\ldots A_s}(X)$ defined on the ambient space $\fR^{d,2}\backslash\{0\}$. Identify them as Taylor coefficients in the power series expansion of a generating function
$\Phi(X,P)\,=\,\sum_s\,\frac1{s!}\, \Phi_{A_1\ldots A_s}(X)\,P^{A_1}\ldots P^{A_s}$ where the $P$'s are mere auxiliary variables. The homogeneity degree in $P$ corresponds to the rank of the tensor field. In addition to the action of $\mathfrak{o}(d,2)$ as
$J_{AB}=L_{AB}+P_{A} \dl{P^{B}}-P_{B} \dl{P^{A}}$ the space of such fields is equipped with an action of $\mathfrak{sp}(4)$ generated by
\begin{equation}
 \begin{gathered}
     S=\dl{X}\cdot\dl{P}\,,\qquad T=\dl{P}\cdot\dl{P}\,,\qquad
\Box=\dl{X}\cdot\dl{X}\,,\\
\bsd=X\cdot\dl{P}\,,\qquad  U_-=P\cdot\dl{P}-X\cdot\dl{X}\,,\qquad S^\dagger=P\cdot\dl{X} \,,\\
\overline{\Box}=X^2,\qquad \bs=X\cdot P\,,\qquad \bar T=P\cdot P,\qquad U_+=P\cdot\dl{P}+X\cdot\dl{X}+d+2\,.\\
\end{gathered}
\label{sp4}
\end{equation} 
There are two obvious automorphisms induced by $P\to -\dl{P}$, $\dl{P}\to P$ or $X\to -\dl{X}$, $\dl{X}\to X$ which will be useful later:
Below we show that ``imposing'' 
the subalgebra formed by the 6 operators
of the first two lines describes non-normalizable Fronsdal fields on $AdS_{d+1}$. Imposing the isomorphic subalgebra obtained after applying the 1st or 2nd automorphism describes respectively the normalizable Fronsdal fields or the conformal currents. Applying then the other (respectively, 2nd or 1st) automorphism describes the shadow fields. These relations show heuristically that they are all intertwined $\mathfrak{o}(d,2)$-modules.

\subsection{Fronsdal fields}

Let us consider the following 6 first-class constraints coming from the first two lines in \eqref{sp4} 
\begin{equation}
\label{Fronsd-const-shad}
     \begin{gathered}
      S=\dl{X}\cdot\dl{P}\,,\qquad T=\dl{P}\cdot\dl{P}\,,\qquad
 \Box=\dl{X}\cdot\dl{X}\,,\\
 \bsd=X\cdot\dl{P}\,,\qquad  U_-=P\cdot\dl{P}-X\cdot\dl{X}\,,\qquad S^\dagger=P\cdot\dl{X} \,,
 		\end{gathered}
\end{equation}
and take as $\algg$ the one-dimensional subalgebra with the generator $\sd$ while the remaining constraints form $\algh$.
Equations of motion and constraints for gauge parameter take the form
\begin{equation}
 S\Phi=T\phi=\Box \Phi=\bsd \Phi=(U_--2)\Phi=0\,, \quad S\lambda=T\lambda=\Box \lambda=\bsd  \lambda=U_-\lambda=0\,.
\end{equation} 
The constraints $\bsd \Phi=0$ and $(U_--2)\Phi=0$ respectively imply that the ambient tensors are tangent to $AdS_{d+1}$ and that the homogeneity degree in $X$ is fixed by the spin. Tangent and homogeneous ambient tensors on the domain $X^2<0$ are in one-to-one correspondence with intrinsic tensor fields on $AdS_{d+1}$, thus the remaining constraints have natural interpretation in terms of AdS tensor fields.
More precisely, the constraints of the first line in \eqref{sp4}, \textit{i.e.} $S\Phi=T\Phi=\Box\Phi=0$, respectively impose the AdS divergencelessness, the tracelessness and the Fronsdal mass-shell whose critical mass defines ``masslessness'' on anti de Sitter spacetime \cite{Fronsdal:1978vb}.
Strictly speaking, one should add a further 7th constraint $(P\cdot\dl{P}-s)\Phi=0$ in order to describe a spin-$s$ Fronsdal field. 

There is an alternative ambient description of the Fronsdal field in terms of the following
subalgebra of $\mathfrak{sp}(4)$:
\begin{equation}
\label{Fronsd-const-curr} 
     \begin{gathered}
     \sd=P\cdot\dl{X}\,,\qquad \bs=X\cdot P\,, \qquad \bar T=P\cdot P\,,\qquad \\
 	\Box=\dl{X}\cdot\dl{X}\,,\qquad S=\dl{X}\cdot\dl{P} \,,\qquad U_+=P\cdot\dl{P}+X\cdot\dl{X}+d+2\,.
 		\end{gathered}
\end{equation}
This can be obtained from \eqref{Fronsd-const-shad} by the transformation $P\to -\dl{P}$, $\dl{P}\to P$. 
The gauge subalgebra $\algg$ is formed by the constraints $\sd\,,\bs\,, \bar T$
while $\algh$ by $\Box, S\,, U_+$.
The equations of motion and the gauge symmetries read explicitly as
\begin{equation}
\Box \Phi=S \Phi=(U_+-4)\Phi=0\,,\qquad \delta_\lambda\Phi=\sd \lambda_1+\bs\lambda_2+\bar T \lambda_3\,.
\end{equation}
To see why these constraints encode Fronsdal fields, one can first solve the homogeneity
constraint for both field and gauge parameters and then employ the gauge transformations generated by $\bar T$ and $\bar S$ to assume $\Phi$ totally traceless and independent of the radial component $X\cdot P$ of the auxiliary variable $P$. This is a rigid gauge fixation as any transformation with $\lambda_1=0$  and nontrivial 
$\lambda^2$ or $\lambda^3$ breaks the condition. 
Furthermore, let us consider the remaining gauge transformation generated by $\sd$.
In order to preserve the gauge condition it can be adjusted by compensating transformations with some $\lambda^{2}$ and $\lambda^{3}$ functions of $\lambda^1$ in order to preserve the tracelessness and the tangency conditions. Using the identification between traceless tangent ambient tensor fields and the traceless AdS tensor fields, one concludes that the gauge transformation is precisely the standard one. Finally, the ambient constraints $\Box \Phi=S \Phi=0$ imply that the respective AdS tensor $\phi$ satisfies the proper mass-shell condition and is divergenceless.

\subsection{Conformal currents}

A rank-$s$ symmetric conformal current $j_{a_1\ldots a_s}(x)$ on conformal space $\manX_d$ is a primary field with weight $\Delta=s+d-2$,
traceless and conserved:
\begin{equation}
\label{current}
 \left(\dl{p}\cdot\dl{p}\right)j(x,p)=0\,, \qquad \left(\dl{x}\cdot\dl{p}\right)j(x,p)=0\,,
\end{equation}
where again tensors $j_{a_1\ldots a_s}(x)$ have been packed into a generating function $j(x,p)$ by making use of an auxiliary variable $p^a$.

The ambient formulation of conformal currents comes from the following
constraints:
\begin{equation}
\label{eq:currents}
        \begin{gathered}
    \bsd =X\cdot\dl{P}\,,\qquad T=\dl{P}\cdot\dl{P}\,,\qquad \overline{\Box}=X^2\,,\\
     S=\dl{X}\cdot\dl{P}\,,\qquad U_+=P\cdot\dl{P}+X\cdot\dl{X}+d+2\,,\qquad  \bar S = X\cdot P\,.
		\end{gathered}
\end{equation}
These constraints can be obtained from \eqref{Fronsd-const-shad} via the transformations $X\to -\dl{X}$, $\dl{X}\to X$ or, equivalently, from
\eqref{Fronsd-const-curr} via the permutation $X\leftrightarrow P$, $\dl{X}\leftrightarrow\dl{P}$.
As a gauge subalgebra $\algg$ we take that of $\overline{\Box}, \bar S$. Note that it is Abelian. The subalgebra
$\algh$ is formed by the remaining constraints $\bsd,T,S,U_+$. In particular equations of motion
take the form
\begin{equation}
 \bsd \Phi=T\Phi=S\Phi=(U_+-4)\Phi=0\,.
\end{equation} 
Concerning the gauge freedom associated with the constraint $\overline{\Box}$, the situation is similar to the scalar field on conformal space except that here homogeneity degree in $X$ leads to the canonical dimension $\Delta=s+d-2$ for the current. The constraint $\bsd\Phi=0$ and the gauge freedom associated to $\bar S$ further imply that the components of the ambient tensor fields are in one-to-one correspondence with the components of an intrinsic tensor fields on the conformal space $\manX_d$. 

This ambient approach to tensor fields in CFTs is by now standard (see \textit{e.g.} \cite{Costa:2011mg} and references therein).
So the remaining constraints find their natural interpretation: the constraints $T\Phi=0$ and $S\Phi=0$ are nothing but the ambient translation of \eqref{current}.
If one relaxes the constraints   
$\bsd\,,T\,,\overline{\Box}$
in \eqref{eq:currents}, then one describes traceful conserved currents on $AdS_{d+1}$ \cite{Bekaert:2010hk}.

\subsection{Shadow fields}

A rank-$s$ symmetric shadow field $\phi_{a_1\ldots a_s}(x)$ on conformal space $\manX_d$ is a primary field with weight $2-s$,
traceless and subject to the Fradkin-Tseytlin gauge transformations \cite{Fradkin:1985am,Segal:2002gd}:
\begin{equation}
\label{shadow}
 \left(\dl{p}\cdot\dl{p}\right)\phi(x,p)=0\,, \qquad \delta_\varepsilon \phi(x,p)=\Pi \left(\,p\cdot\dl{x}\varepsilon(x,p)\,\right)\,,
\end{equation}
where $\Pi$ denotes the projection to the traceless component. The shadow field can be equivalently described as a traceful tensor field with the gauge transformations \cite{Segal:2002gd}:
\begin{equation}
\label{gshadow}
 \delta_\alpha \phi(x,p)= p^2\alpha(x,p)\,, \qquad \delta_\varepsilon \phi(x,p)=\,p\cdot\dl{x}\varepsilon(x,p)\,,
\end{equation}
so that \eqref{shadow} can be seen as the gauge fixing of the Weyl-like gauge transformations in \eqref{gshadow}.

To describe the shadow fields in a manifestly conformal way, let us consider the {following 6 
constraints from~\eqref{sp4}:}
\begin{equation}
\label{amb-shad-const}
 \begin{gathered}
\bsd =X\cdot \dl{P}\,,\qquad U_-=P\cdot \dl{P}-X\cdot \dl{X}\,, \qquad S^\dagger=P\cdot \dl{X}\\
\overline{\Box}=X^2,\qquad {\bar S}= X\cdot P\,,\qquad \bar T=P^2\,.
\end{gathered}
\end{equation} 
These can be obtained from the constraints \eqref{eq:currents} by the transformation $P\to -\dl{P}, \dl{P}\to P$. The gauge subalgebra $\algg$ is formed in this case by the constraints $\sd, \overline{\Box}, \bar S, \bar T$. The remaining constraints of $\algh$ are $\bsd, U_-$. Somewhat as before, the equations of motion 
$\bar S^\dagger \Phi= (U_--2)\Phi=0$ together with the
gauge transformations generated by $\overline{\Box}$ and $\bar S$ allow to restrict to $d$-dimensional tensors
with conformal weight $\Delta=2-s$ as expected for shadow fields.
Consequently, the remaining constraints find their natural interpretation: the gauge freedom associated to the constraints $\bar T$ and $S^\dagger$ are nothing but the ambient version of \eqref{gshadow}.

Note that the constrained system~\eqref{amb-shad-const} has been studied in~\cite{Bekaert:2009fg} where it was shown to describe higher symmetries of the conformal scalar field. In that case, however, the choice of $\algg$
was different. Namely, the gauge transformations were generated by $\overline{\Box}, \bar T,\bar S$ only. Precisely the present choice of $\algg$ for this system was discussed in~\cite{Grigoriev:2006tt} from AdS rather then conformal space perspective. {Note also that the constraints~\eqref{amb-shad-const} can be seen as a linearized constraints 
of a certain nonlinear system~\cite{Bekaert:2009fg,Grigoriev:2006tt} related to the boundary singleton for which $X^A$ and $P_A$ are the ambient space coordinates and momenta.}

\section{Ambient approach to boundary values of AdS fields}
\label{sec:amb-app}

The ambient space
$\fR^{d,2}\backslash\{0\}$ serves {both for fields on $AdS_{d+1}$ and for conformal fields in $d$ dimensions.}
The ambient approach to boundary values consists in two steps: First, one considers a given AdS field $\phi$ and 
reformulates it as an ambient field $\Phi$ of homogeneity degree $-\Delta$ (notice that, normally, for a given AdS field $\phi$ there are two different allowed values for the homogeneity degree of $\Phi$). 
Second, the resulting ambient description is reinterpreted as an ambient description of a conformal field $\phi_0$ with weight $\Delta$ which in turn is identified as a boundary value of the starting point AdS field singled out by the asymptotic behavior $\Delta$. The ambiguity in $\Delta$ results in two different types
of boundary values (\textit{e.g.} for Fronsdal fields: conformal currents and shadow fields).

The identification of the homogeneity degree with minus the conformal dimension can look confusing at first glance as any solution on the hyperboloid can be lifted in the region $X^2<0$ with either homogeneity degree.
The point is that only under the assumption that the ambient field $\Phi$ can be extended consistently over the whole domain $X^2\leqslant 0$, may the conformal field $\phi_0$ be seen as the boundary value of the AdS field $\phi$. 
Indeed, not any AdS field configuration can be lifted to an ambient one on the entire
$\fR^{d,2}\backslash\{0\}$. To see this explicitly, let us for instance, concentrate on the region $X^+\neq 0$.
The evaluation of the ambient field on the null cone reads $\Phi(X^+,-\frac12 X^+x^2,X^+x^a)=(X^+)^{-\Delta}\phi_0(x)$ while its evaluation on
the hyperboloid $X^2=-R^2$ reads as $\Phi(X^+,-\frac12 X^+x^2+\frac{R^2}{2X^+},X^+x^a)=\phi(X^+,x)$. Therefore, the boundary behaviour of the AdS field has to be such that $\lim_{X^{^+}\rightarrow\infty}[(X^+)^{\Delta}\phi(X^+,x)]=\phi_0(x)$.
{One recovers the traditional AdS/CFT formulas by making use of the Poincar\'e coordinates $z=R/X^+$ and $x^a=X^a/X^+$ on the patch $X^+>0$ of the hyperboloid $X^2=-R^2$.}

It is important to mention that here and in the rest of the paper we focus on
the near boundary behavior of AdS fields and hence disregard the behavior in the
interior of the AdS space (\textit{i.e.} the respective region of the manifold
$\amb$ in the ambient terms).  The near-boundary analysis leaves two types of
boundary values unrelated. However, requiring regularity in the interior
determines the current-type  (sub-leading) boundary value in terms of the
shadow-type one and hence allows to obtain the boundary CFT correlation
functions~(for more details see
\textit{e.g.}~\cite{Balasubramanian:1998sn,Skenderis:2002wp} and Refs. therein).
We leave the ambient space implementation of this procedure for the future since
our present concern is near-boundary analysis.

As far as totally symmetric gauge fields are concerned, on the one hand, by
themselves Fronsdal's spin-$s$ equations allow for two possible choices of
boundary behaviour and so describe the direct sum of two indecomposable
$\mathfrak{o}(d,2)$-modules respectively equivalent to the spin-$s$ conformal
current and shadow field. On the other hand, as we have just seen the respective
ambient equations define a single indecomposable $\mathfrak{o}(d,2)$-module if
one considers them on $\amb$ (not only on $X^2<0$) because the fixed homogeneity
degree in $X$ should implement a specific choice of boundary behaviour. More
precisely, constraints~\eqref{Fronsd-const-shad} and \eqref{Fronsd-const-curr}
describe respectively non-normalizable and normalizable Fronsdal fields.

Consider as an illustration, the AdS scalar field. The two different ambient
formulations differ by the choice of the homogeneity degree via
$\Delta_\pm=\frac{d}{2}\pm\sqrt{(\frac{d}{2})^2+(mR)^2}$. For $\Delta=\Delta_+$
constraints~\eqref{adscft} determine the conformal operator of dimension
$\Delta_+$ which is unconstrained (any $\phi_0$ can be extended to the ambient
space such that~\eqref{adscft} are satisfied). For $\Delta=\Delta_-$ the
respective conformal field is to be interpreted as the shadow of the latter
field. If $d-2\Delta \neq 2\ell$ with $\ell$ any positive integer, then any
$\phi_0$ extends to the ambient space field $\phi$
satisfying~\eqref{adscft}. If $d-2\Delta = 2l$ for some positive integer $l$.
Then the extension is obstructed and its existence imposes the conformally
invariant equation $\left(\dl{x}\cdot\dl{x}\right)^{\ell}\phi_0=0$, as will be
explained below. Alternatively one can allow for the logarithmic terms to cancel
the obstruction (see~\cite{Skenderis:2002wp} and references therein for more
details on logarithmic anomalies). Note that the extension is not uniquely
determined by $\phi_0$: the ambiguity is parametrized by $\phi_\ell$ which can
be identified as the scalar conformal current of dimension $\Delta_+$.

Although the identification of the conformal fields associated to the AdS scalar field is relatively straightforward in either the intrinsic AdS or ambient space terms, its extension to more general gauge fields becomes less obvious if one wants to preserve manifest $\mathfrak{o}(d,2)$-invariance and keep track of the gauge symmetries.  
Moreover, the ambient approach is to be generalized in order to allow for locally AdS space 
(\textit{e.g.} ``unfolded'' AdS). In what follows we propose the parent extension of the ambient approach which is more geometrical and is free of the previous drawbacks.

\section{Parent approach to boundary values of AdS fields}
\label{sec:conf-part}

Given an ambient space description of either AdS or conformal gauge field it can be lifted
to the so-called parent formulation which is defined on respectively AdS or conformal space in
generic intrinsic coordinates while the ambient construction is lifted to the target space where it
becomes purely algebraic. This construction was described in details in \cite{Barnich:2006pc}
for AdS Fronsdal fields and then extended to the conformal setting in~\cite{Bekaert:2009fg}
(more general parent formulations and further developments can be found in~\cite{Barnich:2010sw, Grigoriev:2010ic} and \cite{Grigoriev:2006tt,Alkalaev:2008gi,Alkalaev:2009vm}).
The parent formulation is closely related to the unfolded approach~\cite{Vasiliev:1988xc,
Vasiliev:2003ev} and
can also be seen as a (generalization of) Fedosov quantization~\cite{Fedosov:1994} 
of the underlying constrained system. 

\subsection{$\mathfrak{o}(d,2)$ tensor fields on ambient, AdS, and conformal spaces}
\label{confspapar}

Description of the ambient space fields in terms of arbitrary coordinates can be achieved by introducing new variables $Y^A$ interpreted as coordinates on the fibers of the vector bundle
$\bundle{V}(\amb)$ isomorphic to the tangent bundle $T(\amb)$ via a given local frame.
Let $\omega^A_B$, $e^A$ and $V^A$ be respectively an affine connection one-form, a given coframe one-form associated to the invertible map $T(\amb)\to \bundle{V}(\amb)$ and a given section of the bundle
$\bundle{V}(\amb)$ (called ``compensator'') satisfying the standard conditions
\begin{equation}\label{eq:EWV-comp}
  \begin{aligned}
  \omega^C_A\eta_{CB}+\eta_{AC}\omega^C_B=0\,,\qquad &
\derham \omega^B_{A}+\omega^B_C\omega^C_A=0\,,\\
\derham V^A+\omega^A_B V^B=e^A,\qquad &
\derham e^A+\omega^A_B e^B=0\,,
\end{aligned}
\end{equation}
obviously invariant under a change
of local coordinates and local frame, where $\derham$ denotes the de Rham differential on $\amb$.

One next considers an associated vector bundle with the fiber
being the space of formal power series in
$Y$ and polynomials in $P$ variables. Let $\nabla$ be a flat covariant derivative acting in the fiber as follows
\begin{multline}
 \nabla=\derham-e^A\dl{Y^A}- \omega^B_A\Big(Y^A\dl{Y^B}+P^A\dl{P^B}\Big)
=\\
=\derham-\derham V^A\dl{Y^A}-
\omega^B_A\Big(\big(\,Y^A+V^A\,\big)\dl{Y^B}+P^A\dl{P^B}\Big)\,,
 \label{covder}
\end{multline}
where in the second equality we made use of~\eqref{eq:EWV-comp}. The symmetric tensor fields on the ambient 
space can be then identified with the covariantly constant section $\nabla\Psi(X|Y,P)=0$ of this vector bundle.

In order to get back to the ambient description in terms of $\Phi(X,P)$, one needs to take
Cartesian coordinates $X^A$ on $\amb$ and chose the following particular
solution to~\eqref{eq:EWV-comp}: 
\begin{equation}\label{specsol}
\omega^A_B=0,\qquad V^A=X^A,\qquad e^A_B=\delta^A_B\,.
\end{equation}
Vice versa, given a solution to \eqref{eq:EWV-comp} {such that $e$ is invertible} one can (at least locally) find
coordinates $X^A$ and a local frame of $\bundle{V}(\amb)$ such that
\eqref{specsol}.
The covariantly constant sections of \eqref{covder} with \eqref{specsol} indeed have the general form $\Psi(X|Y,P)=\Phi(X+Y,P)$.

The advantage of the description in terms of general coordinates and general
local frame is that it not only simplifies computations by allowing for some
particularly useful frames/coordinates and allows to consider general
locally flat manifolds, but it also appears unavoidable in studying reductions
to spacetime submanifolds (or their quotients). Below we sketch how the
ambient tensor fields can be described in terms of fields on submanifolds.
Details can be found in~\cite{Barnich:2006pc,Bekaert:2009fg} (see also~\cite{Grigoriev:2006tt,Alkalaev:2008gi,Alkalaev:2009vm}).

Given a submanifold of $\manX \subset \amb$ vector bundle $\bundle{V}(\amb)$ can be pulled back to $\manX$. Under the pullback connection $\omega$ and section $V$ induces the connection and the section of $\bundle{V}(\manX)$ satisfying~\eqref{eq:EWV-comp}, where $\derham$ now denotes the de Rham differential of $\manX$.
In particular the covariant derivative \eqref{covder} understood as that on $\bundle{V}(\manX)$
remains flat $\nabla^2=0$.

If one restricts to a neighborhood of $\manX$ in $\amb$ the space of covariantly constant sections of $\bundle{V}(\manX)$ is isomorphic (with a right choice of functional space, though) with that of $\bundle{V}(\amb)$. Indeed, the covariant constancy condition determines a unique extension of a section
on $\manX$ to its neigborhod in $\amb$. In this way field configuration over $\amb$ can be represented
as covariantly constant sections of $\bundle{V}(\manX)$. This important property allows to reformulate
an ambient description in terms of fields defined on $\manX$.

To be more precise let $\manX$ be either $AdS_{d+1}$ or the conformal space $\manX_d$.
The pulled-back connection $\omega_{\mu}^{AB}(x)dx^\mu$ and section $V(x)$  are now 1 and 0-forms on $\manX$ determining the covariant derivative~\eqref{covder}. Note that by construction $e^A_\mu=\nabla_\mu V^A$ has maximal rank.

For later purpose, let us now present a useful local frame $E_A$ of the bundle
$\bundle{V}(\manX)$ in the case where $\manX$ is a conformal space $\manX_d$.{ Namely, 
we take a local frame $E_+,E_-,E_a$ such that (details can be found in~\cite{Bekaert:2009fg}):}
\begin{equation}
\begin{gathered}
 \label{frame}
V^+=1,\quad V^-=0,\quad V^a=0\,, \qquad 
\eta_{+-}=1\,,\quad
\eta_{ab}=diag(-++\ldots+)\,,
\end{gathered}
\end{equation}
and other components of $\eta$ vanish. \textit{i.e.} $E_+,E_-,E_a$ form a light-cone-like basis at each point. 
In particular, $a,b=0,1,\ldots,d-1$ are to be seen as Lorentz indices.
 Taking into account~\eqref{eq:EWV-comp} and adjusting the {embedding} $\manX_d \to \amb$ if necessary one finds
\begin{equation}
\begin{gathered}
\label{conf-con}
\omega_+^a=-\eta^{ab}\omega^-_b=e^a\,,\qquad \omega_-^a=\omega^+_a=\omega^+_+=\omega_-^-=0=e^\pm\,,\\
\omega_a^c \eta^{}_{cb}+\eta^{}_{bc}\omega^c_a=0\,,\qquad
d\omega^a_b+\omega^a_c\omega^c_b=0\,,
\qquad de^a+\omega^a_b e^b=0\,,
\end{gathered}
\end{equation}
so that $\manX_d$ can be  locally seen as Minkowski spacetime provided
one identifies $e_\mu^a$ and $\omega_{\mu a}^b$ as the coefficients of the vielbein
and the connection in the tangent bundle over $\manX_d$. 
 
In the case of $\manX=AdS_{d+1}$ the only difference is that
$V^2=-1$ and a useful local frame reads as:
$V^A=\delta^A_{r}$ and $e^A_\mu=\omega_{\mu \,r}^A$ where $r$ denotes the radial component.
To simplify formulas here and in what follows, we rescale coordinates so that $V^2=-1$ instead of $V^2=-R^2$.

\subsection{Scalar singleton}

As an example, the scalar singleton is briefly reviewed. Let us denote the components of $Y^A$ in the above frame $E_A$ as $y^a=Y^a$, $u=Y^-$, $v=Y^+$. The expressions for the covariant differential \eqref{covder} and the twisted form of the constraints \eqref{sp2} are

\begin{equation}
\begin{gathered}
\label{eq:parent-particle}
  \nabla=\derham-\omega^a_b y^b\dl{y^a}-
e^a (v+1)\dl{y^a}+e^a y_a \dl{u}
\,,\\
Y^2+2u\,,\qquad 
Y\cdot\dl{Y}+\dl{v}+\frac{d+2}{2}\,, \qquad \dl{Y}\cdot \dl{Y}\,.
\end{gathered}
\end{equation}
In the space of formal power series in $Y$, the first constraint in \eqref{eq:parent-particle} is associated to a gauge freedom which allows to get rid of the $u$-dependence in $\psi(x^\mu|u,v,y^a)$ while
the second constraint is imposed (with the shift $d+2\to d-2$ due to the gauge freedom) on the states and then fixes the $v$-dependence in terms of a function $\psi_0(x^\mu|y^a)$ which must be annihilated by $\dl{y}\cdot \dl{y}$ due to the third constraint. The covariant constancy condition is now the unfolded form of the d'Alembert equation on flat spacetime describing a scalar singleton.

\subsection{Boundary values}

Let $T_\alpha, T_i$ denote the $\mathfrak{o}(d,2)$-invariant constraints determining the AdS gauge field
in ambient terms. As before let $T_\alpha$ form a gauge subalgebra $\algg$ and $T_i$ {are} the constraints determining the field equations. The corresponding parent formulation is determined by the following constraints
\begin{equation}
 T^Y_\alpha\,, \qquad T^Y_i,\qquad\nabla
\end{equation} 
{where constraints $T^Y_I$ are obtained by replacing 
$X^A\to Y^A+V^A, \dl{X^A}\to \dl{Y^A}$ and 
$\nabla$ is an AdS version of the covariant derivative~\eqref{covder}.} The condition  $\nabla\Psi=0$ should be treated on an equal footing with $T^Y_i\Psi=0$ so that $\algh$ is formed by $\nabla$ and $T^Y_i$ while $\algg$
by $T^Y_\alpha$. At the same time the ambient field $\Phi(X|P)$ is replaced with the 
$\Psi(x|P,Y)$ defined on $AdS_{d+1}$.

{Note that by construction $\commut{\nabla}{T^Y_I}=0$ and $\commut{T^Y_I}{T^Y_J}=f^K_{IJ}T^Y_K$
with the same structure constants. It follows the parent extension does not
change the ghost terms in the ghost-extended constraints $\hat T^Y_i$. In particular, the structure of the
explicit constraints on fields and gauge parameters is unchanged as well.} That $\nabla$ commutes with e.g.
constraints~\eqref{sp4} can be also seen as follows: if the local frame is chosen such that $V^A=const$
then the covariant derivative~\eqref{covder} can be written as $\nabla=\derham+\omega^{AB}J_{AB}$ 
with $J_{AB}=(Y_A+V_A)\dl{Y^B}+P_A\dl{P^B}-(A \leftrightarrow B)$. It is clear that $J_{AB}$
commutes with $T^Y_I$ because $\mathfrak{o}(d,2)$ and $\mathfrak{sp}\,(4)$ are Howe dual
in this twisted representation space as well. 

Rephrasing in parent terms the ambient approach to boundary values developed in Section~\bref{sec:amb-app}
the prescription amounts to replacing $\Psi(x|P,Y)$ defined on $AdS_{d+1}$ with $\Psi(x|P,Y)$ defined on $\manX_d$, the AdS space covariant derivative $\nabla$ with the conformal one, and the compensator satisfying $V^2=-1$ with the one satisfying $V^2=0$. The resulting parent formulation is by construction manifestly conformal and is equivalent to the ambient space formulation of boundary values. This prescription can be reformulated entirely in the BRST language in which case one treats coordinate differentials $dx^\mu$ as ghost variables and works in terms of the complete BRST operator $\Omega=\nabla+Q$. In this case the above replacement is to be applied to the entire $\Omega$ and $\Psi(x|P,Y,\text{ghosts})$.

The advantage of the parent formulation over the ambient space one is that it explicitly relates the theory
defined on AdS to the theory defined on the boundary. It operates in terms of generic spacetime coordinates
and works equally well for locally AdS spacetimes. As we are going to see next, the parent reformulation has also some technical advantages as it allows to perform computations using special fiber coordinates and at the same time maintain covariance through the use of covariant derivatives. Furthermore this approach extends to the nonlinear level (see Section~\bref{sec:gen} for further details) and has a lot in common with the analogous technique~\cite{Vasiliev:2012vf} in the unfolded framework.

\section{Boundary values of AdS fields}
\label{sec:boundary}

In this section we apply the parent version of the ambient technique to study
boundary values of Fronsdal field on AdS. As a warm-up in Subsection
\bref{sub:AdSscal} we consider Klein--Gordon field on AdS. The analysis of
Fronsdal fields is presented in Subsection \bref{sub:Fronsdal}. 

\subsection{AdS scalar field}\label{sub:AdSscal}
{The parent formulation of a massive scalar field on $AdS_{d+1}$ is known \cite{\BGadS,Grigoriev:2011gp}.} There is no gauge symmetry and the constraints read as
\begin{equation}
 \nabla \Psi(x|Y)=0\,, \qquad  \left((Y+V)\cdot\dl{Y}+\Delta\right)\Psi(x|Y)=0\,, \qquad  \dl{Y}\cdot \dl{Y}\Psi(x|Y)=0\,,
\end{equation} 
where $\nabla$ is AdS version of the covariant derivative \eqref{covder} while $V$ is an AdS version ($V^2=-1$)
of the compensator.

According to the prescription of the previous section the formulation of
boundary values is obtained by a consistent pullback of these structures to the
conformal boundary $\manX_d$. This simply amounts to replacing the AdS
compensator and the covariant derivative by the respective conformal ones and
taking $\Psi$ defined on $\manX_d$.

Concretely, we use the frame~\eqref{frame}, connection \eqref{conf-con} and introduce 
notations $y^a=Y^a$, $u=Y^-$, $v=Y^+$. To start, we notice that any $v$-independent function $\Psi_0(x^\mu|y^a,u)$ can be uniquely extended to a solution 
\begin{equation}\label{psi}
\Psi(x^\mu|y^a,u,v)\,=\,\Psi_0(x^\mu|y^a,u)\,-\,v\,\left(y\cdot\dl{y}+u\dl{u}+\Delta\right)\Psi_0(x^\mu|y^a,u)\,+\,{\cal O}(v^2)
\end{equation}
of 
\begin{equation}
\left(y\cdot\dl{y}+u\dl{u}+(v+1)\dl{v}+\Delta\right)\Psi=0\,.
\end{equation}
The second constraint $\dl{Y}\cdot \dl{Y}=\dl{y}\cdot \dl{y}+2\dl{u}\dl{v}$ imposed on states \eqref{psi} leads to
the condition
\begin{equation}\label{2ndcrt}
\left(\dl{y}\cdot \dl{y}-2\dl{u}\left(y\cdot\dl{y}+u\dl{u}+\Delta\right)\right)\Psi_0=0\,.
\end{equation}
Let us now analyze the dynamical content of the system. To this end we use Cartesian coordinates $x^\mu=x^a$ and local orthonormal frame $e^a_b=\delta^a_b$. The covariant constancy condition reads
$\left(\dl{x^a}-\dl{y^a}+y_a\dl{u}\right)\Psi_0=0$ and any $y$-independent function $\phi(x^a|u)$ can be uniquely extended to a covariantly constant section reading, up to cubic terms in $y$,
\begin{equation}
 \Psi_0(x|y,u)=\left(1+y^a\dl{x^a}+\half y^a y^b \dl{x^a}\dl{x^b}+\half y^2\dl{u}+\ldots\right)\phi(x|u)\,.
\end{equation}
The condition \eqref{2ndcrt} evaluated at $y=0$ then leads to the only equation of motion 
\begin{equation}
\label{scalar-eom}
 \left(\,\Box\,+\,\dl{u}\Big(\,d\,-\,2\big(\Delta+u\dl{u}\big)\,\Big)\,\right)\phi(x|u)\,=\,0\,.
\end{equation} 
where $\Box$ now denotes the $d$-dimensional flat D'Alembertian $\Box=\dl{x^a}\dl{x^a}$.

The dynamical content is clear and can be summarized as follows: \textit{If $\Delta$ is such that 
$d-2\Delta\neq 2\ell$ for some positive integer $\ell$, then any function $\phi_0(x)$ can be uniquely completed to a solution $\phi(x|u)$ of \eqref{scalar-eom}, so that the system describes an off-shell scalar field of conformal dimension $\Delta$ on the boundary. If $d-2\Delta=2\ell$ for some positive integer \sout{nonnegative} $\ell$ then the general solution $\phi(x|u)=\sum_{k=0}\phi_k(x) u^k$ to \eqref{scalar-eom} is parametrized by an on-shell scalar field $\phi_0(x)$ satisfying $\Box^\ell\phi_0=0$ and an off-shell scalar field $\phi_\ell(x)$ of dimension $d-\Delta$.} 
Indeed, if $d-2\Delta=2\ell$ for some  $\ell\in \mathbb N\backslash \{0\}$ then
$\phi_0(x)$ is not anymore unconstrained. Adding, order by order in $u$, terms $\phi_k(x) u^k$ where $\phi_k$ is proportional to $\Box^k\phi_0$ one arrives at the equation $\Box\phi_{\ell-1}=0$ because $d-2(\Delta+\ell)=0$. In this way one finds that $\Box^\ell\phi_0=0$. Furthermore, the system imposes no restrictions on
$\phi_\ell(x)$ entering $\phi(x,u)$ as the term $\phi_\ell(x) u^\ell$. Indeed, it can always be completed by
higher order terms because the coefficient $d-2(\Delta+\ell+k)$ is always nonzero for $k>0$.

The above analysis is algebraically similar to the standard AdS/CFT recipe for obtaining asymptotic solutions as Frobenius series in the radial coordinate (see \textit{e.g.} \cite{Skenderis:2002wp} and references therein). The difference is however that our analysis is purely algebraic and does not require using special coordinate systems on AdS. In fact, it can be performed entirely in the target space.

To anticipate the discussion of the next subsection, let us consider the interesting case of a scalar Fronsdal field on $AdS_{d+1}$ (\textit{i.e.} of mass-square $m^2=2(2-d)R^{-2}$) with boundary prescription corresponding to ``non-normalizable'' modes ($\Delta=2$). For $d$ odd or $d=4$, the boundary data is a scalar shadow field $\phi_0$ but, for even $d\geqslant 6$, the boundary data is encoded in two conformal fields: a scalar Fradkin-Tseytlin field $\phi_0$ of weight $2$ such that $\Box^{\frac{d-4}2}\phi_0=0$ and a scalar ``current'' $\phi_{\frac{d-4}2}$ of weight $d-2$. As one can see, this boundary prescription for the bulk scalar field is unitary for $d\leqslant 6$. Notice that, for any $d\geqslant3$, the normalizable boundary prescription corresponds to $\Delta=d-2$ and is encoded into a single conformal ``current'' $\phi_0$, in agreement with the irreducibility of the conformal current modules.

\subsection{Fronsdal field}\label{sub:Fronsdal}

The twisted form of the constraints \eqref{Fronsd-const-shad} read as
\begin{equation}
\label{eq:operators}
    \begin{gathered}
     S=\dl{Y}\cdot\dl{P}\,,\qquad S^\dagger=P\cdot\dl{Y} \,,\qquad T=\dl{P}\cdot\dl{P}\,,\qquad
\Box_Y=\dl{Y}\cdot\dl{Y}\,,\\
\bsd=(Y+V)\cdot\dl{P}\,,\qquad  U_-=P\cdot\dl{P}-(Y+V)\cdot\dl{Y}\,.
\end{gathered}
\end{equation}
To begin with, let us disregard gauge invariance and only investigate the allowed field configurations, \textit{i.e.} solutions $\Psi(x|Y,P)$ of 
\begin{equation}
\label{constr}
\begin{gathered}
\nabla \Psi=0\,,\quad (U_--2)\Psi=0\,, \quad  \bsd \Psi=0\,,\\
\Box_Y\Psi=0\,,\quad S\Psi=0\,, \quad T \Psi=0\,.
\end{gathered}
\end{equation} 
Let us concentrate first on the equations in the first line of~\eqref{constr} and introduce the notation $p^a=P^a$, $w=P^-$, $w^\prime=P^+$. In the adapted frame and coordinates used in the previous subsection, one explicitly has
\begin{align}
 \left(\dl{x^a}-(v+1)\dl{y^a}+y_a\dl{u}-w^\prime\dl{p^a}+p_a\dl{w}\right)\Psi=0\,,\\
 \left(y\cdot\dl{p}+u\dl{w}+(v+1)\dl{w^\prime}\right)\Psi=0\,,\\
 \left(y\cdot\dl{y}+u\dl{u}+(v+1)\dl{v}-P^A\dl{P^A}+2\right)\Psi=0\,.
\end{align} 
It is easy to find the general solution to these equations. Indeed any function $\phi(x^a|u,w,p^b)$ can be extended to $\Psi(x^a|Y^A,P^B)$ satisfying these three equations and such that
$\Psi|_{y=v=w^\prime=0}=\phi$. Indeed, there are terms $\dl{y^a}$, $\dl{w^\prime}$, and $\dl{v}$
in respectively the first, second, and third equation which can be used to construct the solution order by order in $y^a,w^\prime,v$. 
Therefore it is possible to rewrite the constraints of the second line
of~\eqref{constr} solely in terms of $\phi$. 
Moreover, in order to simplify those equations, let us restrict to a particular spin: $(p\cdot\dl{p}+w\dl{w}-s)\phi=0$. 
The constraints of the second line of~\eqref{constr} now become
\begin{align}
\tilde\Box \phi+\dl{u}\left(d+2(s-2)-2u\dl{u}\right)\phi=&0\,,\label{eq1}\\
\left(\dl{p}\cdot\dl{x}\right)\phi +\dl{w}\left(d+1+2(s-2)-w\dl{w}-2u\dl{u}\right)\phi=&0\,,\label{eq2}\\
\left(\dl{p}\cdot\dl{p}\right)\phi-2u\left(\dl{w}\right)^2\phi=&0 \label{eq3}\,,
\end{align}
where $\tilde\Box=\Box+2(p\cdot\dl{x})\dl{w}+p^2(\dl{w})^2$.

In terms of $\phi$ the gauge transformation read as
\begin{equation}
\label{phi-gt}
 \delta\phi=\left(p\cdot\dl{x}+p^2\dl{w}+w\dl{u}\right)\lambda\,,
\end{equation}
where the gauge parameter $\lambda(x|u,w,p)$ must obey to the analogue of equations \eqref{eq1}-\eqref{eq3}
obtained by replacing $\phi$ with $\lambda$ and $s-2$ with $s-1$. In other words, the gauge parameter obeys to differential constraints which is the price to pay for our partial gauge-fixing.
As in the scalar case, the system described by~\eqref{eq1}--\eqref{eq3} and the residual gauge symmetries \eqref{phi-gt} are drastically different in odd and even dimensions.

\subsubsection{Odd boundary dimension -- Shadow field:}

To begin with let us concentrate on the case where $d$ is odd. In this case,
the operator $d+2(s-2)-2u\dl{u}$ in \eqref{eq1} has no zero eigenvector in a space of power series in $u$, so that there is no obstruction in solving the first equation order by order in $u$. 
In~\bref{app:odd}, we show
that any field $\phi_0^0(x|p)$ 
(recall that $p^a\dl{p^a}\phi^0_0=s\phi^0_0$ as we are describing spin-$s$ field)
satisfying $(\dl{p}\cdot\dl{p})\phi_0^0=0$ can be completed to a solution $\phi(x|u,w,p)$ 
satisfying~\eqref{eq1}--\eqref{eq3} and such that $\phi|_{u=w=0}=\phi_0^0$. The proof is purely technical
and straightforward.

Analyzing the gauge invariance in a similar way, one finds that \textit{for $d$ odd, the general solution to the system~\eqref{eq1}-\eqref{eq3} modulo the residual gauge symmetries \eqref{phi-gt} is parametrized by a shadow field $\phi(x,p):=\phi(x|u=0,w=0,p)$ defined by \eqref{shadow}.}

A remarkable manifestation of this fact is that in the ambient space description
the two systems of constraints related by $P\to -\dl{P}$ and $X \to -\dl{X}$ are equivalent in odd dimensions
but (as we are going to see next) are not equivalent in even dimensions.

\subsubsection{Even boundary dimension -- Fradkin--Tseytlin field \& conformal current:}

\label{sec:even-d}

If the dimension is even the coefficient in~\eqref{eq1} vanishes for a certain power of $u$ which gives rise to
constraints on $\phi_0$ as in Subsection \bref{sub:AdSscal}. 

More precisely, repeating the analysis of the previous section one finds
that any $\phi^0_0(x|p)$ can be uniquely extended to $\phi_0(x|p,w)$ satisfying~\eqref{eq2}. 
Let $\ell=\frac{d-4}{2}+s$ denote a positive integer since by assumption $d\geqslant 4$ and $s>0$. Assuming that $\phi^0_0$ is
annihilated by $\dl{p}\cdot\dl{p}$ allows to iteratively construct $\phi=\phi_0+u\phi_1+\ldots+\frac{1}{(\ell-1)!}\phi_{\ell-1}$ up to order $\ell-1$ such that equations \eqref{eq1}--\eqref{eq3} are fulfilled to this order in $u$. At the next order
one obtains $\tilde\Box \phi_{\ell-1}=0$ so that 
\begin{equation}
\label{FT+pgc}
 \tilde\Box^\ell
 \phi_0=0
 \,,\quad 
 \left(\dl{p}\cdot\dl{x}\right) \phi_0+
 \dl{w}(2\ell+1-w\dl{w})\phi_0=0\,, \quad
 \left(\dl{p}\cdot\dl{p}\right)\phi_0=0\,.
\end{equation} 
where for convenience we have also explicitly listed the equation defining $\phi_0$
in terms of $\phi_0^0$ and the trace constraint.
Recall that the second equation uniquely determines $\phi_0(x|p,w)$ in terms of the traceless $\phi_0^0(x|p)$.

It is easy to see that equation~\eqref{FT+pgc} is gauge invariant under the gauge transformation \eqref{phi-gt}
provided $\lambda(x|p,w,u)$ satisfy the gauge parameter version of the equations \eqref{eq1}--\eqref{eq3},
in particular $\tilde\Box^{\frac{d-2}{2}+s}\lambda_0=0$,
where $\lambda_0=\lambda|_{u=0}$. More precisely, the variation of the first equation in \eqref{FT+pgc}
under \eqref{phi-gt} has the form
\begin{equation}
\label{gt-cond}
 \tilde\Box^{\frac{d-4}{2}+s}\left(p\cdot\dl{x}+p^2\dl{w}+w\dl{u}\right)\lambda_0\,=\,
\,-\,\frac{2}{d-4+2s}\, w\, \tilde\Box^{\frac{d-2}{2}+s}\lambda_0\,,
\end{equation} 
where we have assumed that $\lambda$ satisfies \eqref{eq1} (with $s-2\to s-1$) up to order $u^{\frac{d-2}{2}+s}$
and made use of the relation $\commut{\tilde\Box}{w}=2(p^a\d_a+p^2\dl{w})$\,. Here by slight abuse of notations $\dl{u}\lambda_0$ stand for $(\dl{u}\lambda)|_{u=0}$.

Let us consider the following obvious consequence of~\eqref{FT+pgc}:
\begin{equation}
\label{FT}
 (\tilde\Box^{\frac{d-4}{2}+s}\phi_0)|_{w=0}=0\,.
\end{equation} 
Because the gauge transformation of \eqref{FT+pgc} is proportional to $w$ the above equation
is gauge invariant even if $\lambda$ satisfies \eqref{eq1} (with $s-2\to s-1$) up to order $u^{\frac{d-4}{2}+s+1}$ only.
This allows to take $\lambda_0^0(x|p)$ unconstrained (except from the traceless condition $\dl{p}\cdot\dl{p}\lambda^0_0=0$). By inspecting
the gauge transformation in terms of $\phi_0^0$ one finds the usual transformation law with unconstrained gauge
parameter. Taking into account the order in the derivatives, conformal invariance, {and known classification statements~\cite{Shaynkman:2004vu}} one concludes that~\eqref{FT} imposes the Fradkin--Tseytlin equations on~$\phi_0^0(x,p)$. Those equations in~\eqref{FT+pgc} which are not contained in~\eqref{FT} are 
to be interpreted as partial gauge conditions originating from the very first partial gauge fixation we have made.

Leaving the technical details of the derivation
for~\bref{app:even}
let us summarize the boundary data: in addition to the Fradkin-Tseytlin field $\phi_0^0$ the system simultaneously describes the conformal current $\psi_\ell^0$. Although in this formulation $\psi_\ell^0$ does transform under gauge transformations, it is not a genuine gauge field because if one gauge fixes the Fradkin-Tseytlin field then this gauge symmetry is not present anymore (just like usual gauge symmetry for matter fields in Yang-Mills theory).

Let us summarize this result more formally: \textit{For $d$ even, the general solution to the system~\eqref{eq1}-\eqref{eq3} is parametrized by a Fradkin-Tseytlin field $\phi^0_0(x,p):=\phi(x|u=0,w=0,p)$
satisfying the equations of motion encoded in~\eqref{FT+pgc} and by a conformal current $j(x|p):=(\dl{u})^{\frac{d-4}2+s}\phi|_{u=w=0}$ satisfying \eqref{current}, the remaining components
are either uniquely determined through~\eqref{eq1}-\eqref{eq3}
or pure gauge.} The structure of the gauge symmetries is as follows: \textit{The equations~\eqref{FT+pgc} contain the equations \eqref{FT} which are equivalent to the Fradkin--Tseytlin equations on $\phi^0_0=\phi_0|_{u=0}$ invariant under the gauge transformations with traceless parameter $\lambda^0_0(x,p):=\lambda(x|u=0,w=0,p)$, but the entire system of equations of motion encoded in~\eqref{FT+pgc} is only invariant under the gauge transformations with $\lambda^0_0(x,p)$ satisfying the constraints encoded in the gauge parameter version of~\eqref{FT+pgc}.}

\subsubsection{Normalizable solutions} 
According to the general discussion to obtain the boundary values corresponding to the current asymptotic
we need to start with the ambient formulation based on~\eqref{Fronsd-const-curr}. In this case the analysis {simplifies because $(U_+ +d-2)\Psi=0$} uniquely determines the $v$-dependence in terms of $\Psi|_{v=0}$.
This happens because the respective coefficient never vanishes in contrast to the case of constraint $U_--2$. Using this observation one finds that gauge transformation generated by $\sd$ allows to eliminate $w^\prime$, the one generated by $\bar S$ eliminates $w$, and $\bar T$ allows to take the field traceless. Finally, $\Box_Y\Psi=0$ uniquely fixes the dependence on $u$ while the remaining constraint $S$ reproduces the current conservation condition.

To conclude the discussion of totally symmetric fields, we mention that the
standard approach to boundary values of gauge fields is based on using some
gauge-fixing condition where equations of motion reduce to Klein--Gordon
equations with the specific mass-like term (note
however~\cite{Metsaev:2008fs,Metsaev:2009ym}). In contrast, one advantage of our
approach is gauge covariance, which should be useful in higher-spin holography
due to the important role played by gauge symmetries in this context.
Furthermore, the boundary values are studied for both gauge fields and gauge
parameters.\footnote{Strictly speaking, till now we actually employ a somewhat
simplified setting which requires partial gauge fixation which is, actually, a
purely technical simplification (see the discussion in the next section).} 
Another advantage is the manifest $o(d,2)$ covariance of the construction which
guarantees that the choice of asymptotic behavior is not only compatible with
the gauge symmetry but is also $o(d,2)$-covariant.

\section{Generalization}\label{sec:gen}

We now sketch how the approach pushed forward in this paper extends to general gauge systems.
Suppose we are given with a gauge theory defined on a spacetime manifold $\manX$. 
In the BRST language, fields of the theory $\Psi^\alpha$ also include ghost fields and antifields and the theory is determined by the BRST differential $s$. More precisely, $s$ is defined on the jet space -- i.e the space with coordinates $\Psi^\alpha$, their spacetime derivatives $\Psi^\alpha_{A\ldots}$, and the spacetime coordinates $V^A$ and their differentials $e^A\equiv dV^A$ treated as ghost variables. The BRST operator $s$ is nilpotent, carries ghost degree $1$ and commutes with the total derivative $\d^T_A$ (see e.g.~\cite{Barnich:2000zw} for more details on jet space BV-BRST formalism). 

Following the procedure of~\cite{\BGnlp} (see also references therein) the parametrized parent BRST formulation of the system is constructed as an AKSZ sigma model with the target space being the jet space equipped with the differential $Q=-d_H+s$, where $d_H=e^A\d^T_A$, and the source space being $\manX$ extended by the de Rham differentials $\Theta^{\ul{A}}=dX^{\ul{A}}$ seen as Grassmann odd variables of ghost degree $1$.  Here $X^{\ul{A}}$ denote generic coordinates on $\manX$.  In particular, all the variables $V^A,e^A,\Psi^{\alpha}_{A\ldots}$
become fields depending on $X^{\ul A},\Theta^{\ul A}$. The sigma-model equations of motion and gauge symmetries for fields $V^A(X)$ and $e_{\ul A}^B(X)$ read as
\begin{equation}
\label{Z-eq}
 \d_{\ul B}V^A-e^A_{\ul B}=0\,, \qquad \delta_\lambda V^A=\lambda^A\,, \quad \delta_\lambda e^A_{\ul B}=\d_{\ul B} \lambda^A \,.
\end{equation} 
Here and below we use the same notations for the target space coordinate and its
ghost degree zero component field, e.g. $e^A_{\ul{B}}$ denotes linear in
$\Theta^{\ul{B}}$ components of $e^A(X,\Theta)$. In the gauge $V^A(X)=X^A$,
where $X^A$ are suitable coordinates on $\manX$, one has $e^A_B=\delta^A_B$ and
the formulation reduces to a non-parameterized parent BRST
formulation~\cite{Barnich:2010sw} of the starting point system. 

Note that the manifold $\manX$ enters the sigma model in two different roles: as
a part of the target space and as the source space. In general one can replace
the source space with a different manifold. Indeed, at the level of equations of
motion the source manifold is an independent data for an AKSZ sigma model so
that one can consider a family of models with various space-time manifolds but
one and the same target space.

In particular, taking as source a submanifold $\manX_0\subset \manX$ results in
a gauge theory defined on $\manX_0$.
Moreover, if $e^A_{\ul{B}}(X)$ and $V^A(X)$ is a particular background solution
(this can also be understood as the choice of gauge) of the sigma model on
$\manX$ their pullback to $\manX_0\subset \manX$ define a background for the
model on $\manX_0$. This gives a systematic way to identify a gauge theory on
$\manX_0 \subset \manX$ induced by the one on $\manX$.

One can give an alternative interpretation to the choice of source manifolds for
a fixed target space. As we have seen above if the source space is $\manX$
itself the above AKSZ sigma model is equivalent to the starting point system
provided the allowed field configurations are such that
$\rank{(e^A_{\ul{B}})}=\dim{\manX}$ (e.g. gauge $V^A=X^A$ is admissible). Note
that although this condition does not restrict infinitesimal gauge
transformations, finite ones are in general restricted. Besides this natural
phase (where $\rank{e^A_{\ul{B}}}=\dim{\manX}$) one can consider other phases of
the theory. In particular, if $\rank{e^A_{\ul{B}}}=k$ with $k< \dim{\manX}$
functions $V^A(X)$ can be seen as defining a map from the space-time to a
submanifold $\manX_0\subset \manX$. In this case it is natural to take the
spacetime manifold $k$-dimensional because anyway the coordinates along the zero
vectors of $e^A_{\ul{B}}$ are essentially passive (we systematically disregard
subtleties related to global geometry) so that the system effectively lives on
$\manX_0$. This phenomenon is well-known in the context of parametrized
Hamiltonian systems (see \textit{e.g.} the discussion
in~\cite{Henneaux:1992ig}). In this case, in addition to the ``natural'' gauges
where $t=T(\tau)$ is invertible one can consider ``frozen evolution'' gauge
where $t=\const$ and the rank of $e$ vanishes.

To make contact with the approach in the previous sections let us take $\manX=\amb$ and assume in addition that
the starting point theory is $\mathfrak{o}(d,2)$-invariant. More precisely, if $K_{AB}=-K_{BA}$ are $\mathfrak{o}(d,2)$ parameters the transformation of jet-space coordinates reads as
\begin{equation}
\begin{gathered}
JV^A=K^A_B V^B,\quad
J e^A=K^A_B e^B,\quad
J \Psi^\alpha(Y)= K_{A}^{B}[-Y^A \dl{Y_{B}}  \delta^\alpha_\beta+(M_B^A)^\alpha_\beta]\Psi^\beta(Y)
\end{gathered}
\end{equation} 
where we have assumed symmetry is realized on fields linearly and in the last equation we
employ generating function $\Psi^\alpha(Y)=\Psi^\alpha+\Psi^\alpha_A Y^A+\half
Y^A Y^B \Psi^\alpha_{AB}+\ldots$ for $\Psi^\alpha_{A\ldots}$. Direct computations show $\commut{d_H}{J}=0$.  In this language $\mathfrak{o}(d,2)$ invariance of the system is expressed by
$\commut{s}{J}=0$ as we assume in what follows. 
Given such $J$ one can promote the parameters $K^{AB}$ to Grassmann odd 
ghost coordinates $\omega^{AB}$ and extend $J$ such that $J\omega^A_B=\omega^A_C\omega^C_B$. In other
words $J$ becomes a cohomology differential of $\mathfrak{o}(d,2)$ with coefficients in local functions of $\Psi^\alpha_{A\ldots},V^A,e^A$.

Now we can repeat the construction of this section with $Q=-d_H+s$ replaced with $Q^\prime=-d_H+J+s$.
The AKSZ sigma model equations of motion for fields $V^A,\omega^{AB},e^A$ where $\omega^{AB}$ denotes a $1$-form component of $\omega^{AB}$ are precisely equations~\eqref{eq:EWV-comp}.  Let us stress that fields $V^A$ originating from the starting point  spacetime coordinates entering the formulation through  the parametrization are now interpreted as components of the compensator field.

If one applies the procedure outlined above to the ambient systems considered in the preceding sections 
and then replaces $\manX$ with either conformal space $\manX_d$ or $AdS_{d+1}$ one reproduces
the respective parent formulations and the relation between the AdS system and its associated boundary system. 
It is instructive to illustrate, how the covariant derivative arises automatically once a particular background solution for $V^A$ is taken. Restricting for definiteness to the AdS case so that $V^2=-1$ and $V,e,\omega$ satisfy the AdS version of~\eqref{eq:EWV-comp}, the expression for the total BRST differential $s^P=\derham-d_H+s+J$ becomes
\begin{equation}
s^P\Psi^\alpha=\nabla \Psi^\alpha + s \Psi^\alpha\,, \qquad 
\nabla \Psi^\alpha =\derham \Psi^\alpha-\omega^A_B Y^B\dl{Y^A}\Psi^\alpha-e^A\dl{Y^A}+\omega^A_B (M^B_A)^\alpha_\beta \Psi^\beta
\end{equation} 
so that $\nabla$ coincides with~\eqref{covder} if one uses the gauge where $V^A$ is constant.

In a certain sense all the three space-time realizations (ambient, AdS, conformal) of the background independent AKSZ sigma model can be considered equivalent (in particular their local BRST cohomology groups are directly related~\cite{Barnich:2009jy,Barnich:2010sw}). When formulated in these terms our approach becomes very similar to the unfolded approach to boundary dynamics~\cite{Vasiliev:2012vf}. Indeed, AKSZ sigma model can be seen as a Batalin-Vilkoviski-BRST extension of a free differential algebra with constraints (see \textit{e.g.}~\cite{Barnich:2005ru}). However, an important extra ingredient is the presence of the compensator field $V^A$ whose inequivalent vacuum solutions distinguish different phases.  In particular, the action of $\mathfrak{o}(d,2)$ on fields
depends on the choice of the vacuum solution for $V^A$ because the twisted realization of the $\mathfrak{o}(d,2)$ (local) action involves $V^A$.

\section{Conclusions}

In this paper, we have developed the ambient space approach to boundary values of AdS gauge fields. Starting from an ambient formulation of a given $AdS$ field one reinterprets it as an ambient formulation of a certain conformal field which is then identified  as the boundary value of the original $AdS$ field, with the asymptotic behavior determined by the choice of homogeneity degree in the ambient formulation. 

This procedure can be seen as a map that sends a gauge theory on $AdS_{d+1}$ to a conformal (gauge) theory in $d$ dimensions.
For a generic Fronsdal field, the boundary data is encoded in two distinct conformal systems -- the one describing boundary values of the normalizable solutions (\textit{i.e.} the conformal currents) and the one for non-normalizable solutions (\textit{i.e.} the shadow fields). However, for a Fronsdal field in odd-dimensional AdS space (and hence even $d$) the boundary theory for non-normalizable solutions simultaneously describes both  
the conformal current and the Fradkin-Tseytlin field.\footnote{Notice that, from the group-theoretical viewpoint, in both case{s} the non-normalizable Fronsdal field is parametrized by the shadow field module as it should, but in even dimension $d$ this $\mathfrak{o}(d,2)$-module is seen as the semidirect sum of the Fradkin-Tseytlin module and the conformal current submodule.} We stress that this Fradkin-Tseytlin field is an \textit{on-shell} shadow field in the sense that it is subject to Fradkin-Tseytlin equations~\cite{Fradkin:1985am,Segal:2002gd},
which naturally arise here in a generating formulation somewhat similar to that proposed in~\cite{Metsaev:2009ym}. This can be traced to the fact that in our approach we use the unfolded-type technique
which in the minimal version does not allow for logarithmic terms~\cite{Skenderis:2002wp} to cancel the obstruction.
We also discussed how this approach extends to more general setting and discuss its relation to the unfolded technique of~\cite{Vasiliev:2012vf}.

We expect that thanks to the similarity with the unfolded framework the approach
can be useful in extending the considerations of~\cite{Vasiliev:2012vf} to
nonlinear higher spin gauge theories to $AdS_{d+1}$ with $d>3$. In particular,
it is natural to expect that for even $d$ the on-shell shadow field and the
conformal current both enter the nonlinear theory of boundary values. This is
supported by the structure of gauge transformations for these fields. Indeed, in
this case both of them are affected by the gauge transformation so that, at
nonlinear level where the gauge parameters take values in the higher spin
algebra~\cite{Vasiliev:2003ev}, the theory of boundary values should involve
both fields for all spins in a nonlinear way.

\section*{Acknowledgments}

We are grateful to G.~Barnich for collaboration at the early stage of this project. 
We also wish to thank N.~Boulanger, C.~Iazeolla, E.~Joung, E.~Meunier, D.~Ponomarev, E.~Skvortsov, P.~Sundell, M.~Vasiliev, A.~Waldron and, especially, K.~Alkalaev and R.~Metsaev for useful discussions.
We acknowledge the Schr\"odinger Institute (Vienna) for hospitality where this work was partially performed.

\appendix

\section{Details of the analysis -- odd $d$}
\label{app:odd}
Here we prove that any $\phi_0^0(x|p)$ 
satisfying $(\dl{p}\cdot\dl{p})\phi_0^0=0$ can be completed to a solution $\phi(x|u,w,p)$ 
satisfying~\eqref{eq1}--\eqref{eq3} and such that $\phi|_{u=w=0}=\phi_0^0$. 

To this end, we first show that
there exists a unique field $\phi_0(x|w,p)$ satisfying~\eqref{eq2} and such that $\phi_0|_{w=0}=\phi^0_0$. Indeed \eqref{eq2} determines a unique lift $\phi_0$ as a power series in $w$ provided the operator $d+1+2(s-2)-w\dl{w}$ has no zero eigenvalues on $\phi_0$ (which is independent of $u$). For a homogeneous term of order $k>0$ in $w$, the vanishing of the coefficient would imply $d-3+2s-k=0$ which, for $s>0$ and $d>2$ 
(recall that this equation is absent if $s=0$ while $d>2$ by assumption), never happens because
$k\leq s$ as $(p\cdot\dl{p}+w\dl{w}-s)\phi=0$.

Starting with $\phi_0(x|w,p)$ one finds a unique solution $\phi(x|u,w,p)$ to~\eqref{eq1} such that $\phi|_{u=0}=\phi_0$.
This is possible as the respective coefficient in ~\eqref{eq1} is non-vanishing when $d$ is odd. Furthermore,
$\phi$ satisfies equation~\eqref{eq2} as well. Indeed, decomposing $\phi$ as 
$\phi=\sum_{{l=0}} \frac{1}{l!}u^l \phi_l$
equation \eqref{eq1} says 
\begin{equation}
\label{exp-k+1}
 \phi_{k+1}=-\frac{1}{d-6+2s-2k}\tilde\Box \phi_{k}\,,
\end{equation} 
Suppose that \eqref{eq2} is fulfilled for all $\phi_l$ with $l\leq k$.
Substituting above $\phi_{k+1}$ to \eqref{eq2} one gets
\begin{equation}
\left[\,\left(\dl{p}\cdot\dl{x}\right) \,+\,\dl{w}\left(d-3+2s-w\dl{w}-2(k+1)\right)\,\right]\tilde\Box \phi_k=0\,.
\end{equation} 
Using then
\begin{equation}
 \begin{aligned}
\commut{\dl{p}\cdot\dl{x}}{\tilde\Box}&=2(\d^a\d_a+p^a\d_a\dl{w})\dl{w}\,,\\
\commut{-\dl{w}w\dl{w}}{\tilde\Box}&=
2\left({p^a}\d_a \dl{w}+ p^2 \frac{\d^2}{(\d w)^2}\right)\dl{w}
\end{aligned}
\end{equation} 
along with $(p\cdot\dl{p}+w\dl{w}-s)\phi_l=0$ one finds that~\eqref{eq2} is fulfilled provided \eqref{eq1} does.
Finally, analogous considerations show that the field $\phi(x|p,u,w)$ obtained from $\phi_0^0(x|p)$
such that $(\dl{p}\cdot\dl{p})\phi^0_0=0$ satisfies \eqref{eq3}. 

\section{Details of the analysis -- even $d$}

\label{app:even}

In the case of even  $d$ the coefficient in \eqref{eq1} vanishes for a term  $u^{\ell}\phi_\ell$ with $\ell=(d-4)/2+s$ so that as we have seen in the main text $\phi^0_0$ is subject to equations encoded in~\eqref{FT+pgc}. Moreover, $\phi_\ell$ is not determined through $\phi_0^0$ by \eqref{eq1}. However, \eqref{eq3} does determine $(\dl{p}\cdot\dl{p})\phi_\ell$
in terms of $\phi_{\ell-1}$ and hence in terms of $\phi_0^0$. More precisely, one can show that given $\phi_{\ell-1}$ satisfying~\eqref{eq2} one can construct $\phi_\ell$ satisfying \eqref{eq2} and such that
$(\dl{p}\cdot\dl{p})\phi_\ell=2\ell \frac{\d^2}{(\d w)^2}\phi_{\ell-1}$ (i.e.~\eqref{eq3} is fulfilled
at this order). To this end observe that $(\dl{p}\cdot \dl{x}) \phi^2_{\ell-1}=0$ (recall the expansion
$\phi_k=\sum_{m=0}\frac{1}{m!}\phi_k^m w^m$) thanks to \eqref{eq2}. We then take a $\phi_\ell^0$ such that
$(\dl{p}\cdot\dl{p})\phi_\ell^0=2\ell\phi_{\ell-1}^2$ and $(\dl{p}\cdot \dl{x}) \phi_\ell^0=0$, which exists from
the standard results on the structure of polynomials in $2d$ variables (see e.g.~\cite{\BGST}).

Furthermore we take $\phi_\ell^1$ such that $(\dl{p}\cdot\dl{p})\phi^\ell_1=2\ell\phi_{\ell-1}^3$. Note that equation \eqref{eq2} imposes no constraints
on $\phi^\ell_1$ and $\phi^{\ell-1}_3$. Given $\phi^\ell_0, \phi_\ell^1$ equation \eqref{eq2} uniquely
determines all the $\phi_\ell^l$ with $l>1$. It is then a matter of direct computation to
show $(\dl{p}\cdot\dl{p})\phi^l_\ell=2\ell\phi^{l+2}_{\ell-1}$ so that \eqref{eq3} is satisfied at this order. Taking constructed $\phi_\ell$
as a boundary condition one then finds all the higher order terms $\phi_{l}$ with $l>\ell$ using \eqref{eq2}
(note that the coefficient never vanishes in this case so that the solution exists and is unique).
In this way we completed the proof that any $\phi^0_0$ can be lifted to a solution of \eqref{eq1}--\eqref{eq3} 
provided $\phi^0_0$ satisfies the consistency equations encoded in \eqref{FT+pgc}.

Let us now turn to the solutions to~\eqref{eq1}-\eqref{eq3} which are not determined in terms of $\phi^0_0(x|p)$.
This arbitrariness is described by a traceless element $\psi_\ell(x|p,w)$ satisfying $\eqref{eq2}$ which can be added to $\phi_\ell(x|p,w)$ without spoiling the lower order equations. Just like $\phi_0(x|p,w)$, such a $\psi_\ell(x|p,w)$ can be lifted to a unique solution $\psi(x|p,w,u)$ of~\eqref{eq1}-\eqref{eq3} such that 
$\psi=\frac{1}{\ell!}u^\ell\psi_\ell+O(u^{\ell+1})$ provided $\psi_\ell$ satisfies certain constraints identified shortly. This simply follows by observing that the coefficients in~\eqref{eq1} are always non vanishing for those higher order terms in $u$. To describe this general solution $\psi$, there remains to analyze equation \eqref{eq2} imposed on $u^\ell\psi_\ell(x|p,w)$ and its gauge symmetries (without forgetting that $(\dl{p_a}\dl{p^a})\psi_\ell=0$).

Actually, Equation~\eqref{eq2} immediately implies $(\dl{p}\cdot\dl{x})
\psi_\ell^0=0$, where we used the notation $\psi_\ell=\sum_{l=0}
\frac{1}{l!}w^l\psi_\ell^l(x,p)$, because
$(d-3+2s-w\dl{w}-2u\dl{u})(w\,u^\ell\,\psi^1_\ell)=0$. In this way one finds the
current conservation condition for $\psi_\ell^0$.  In addition one finds that
$\psi^1_\ell$ is undetermined by both~\eqref{eq1} and \eqref{eq2}. At the same
time, the higher components $\psi_\ell^{l}$ with $l\geq 2$ are uniquely
determined by~\eqref{eq2} in terms of $\psi_\ell^1$. 

We now analyze the gauge invariance. Let us consider first the gauge
transformation~\eqref{phi-gt}. For $\psi_\ell^0$ the contribution
$w\dl{u}\lambda$ is not present. The remaining contribution involves
$\lambda_\ell$ only which is uniquely determined by $\lambda_0^0$. As we have
seen $\lambda_0^0$ is naturally interpreted as gauge parameter for FT field
$\phi_0^0$ and should not be taken into account when describing inequivalent
configurations for $\psi_\ell^0$. Indeed, if we impose on $\phi_0^0$ a gauge
condition that removes the gauge freedom completely, then $\lambda_0^0$ vanishes
and hence $\lambda_\ell$ also, so that this gauge symmetry for $\psi_\ell^0$
should be disregarded. 

As far as $\psi_\ell^1$ is concerned the gauge transformation for it contains
the contribution $w\dl{u}\lambda$ which shifts $\psi_\ell^1$ by
$\lambda_{\ell+1}^0$. In its turn $\lambda_{\ell+1}^0$ is an independent gauge
parameter (independent of $\lambda^0_0$) and can be used to gauge away that
component of $\psi_\ell^1$ which satisfies $(\dl{p}\cdot \dl{x}) \psi_\ell^1=0$
because \eqref{eq2} implies $(\dl{p}\cdot\dl{x})\lambda_{\ell+1}^0=0$. Recall
that just like for $\phi_\ell$ only the traceless part of $\lambda_{\ell+1}$ is
independent of $\lambda_0$. Furthermore, gauge parameter $\lambda_{\ell+1}^1$ is
not constrained by \eqref{eq2} and hence its traceless part is free and can be
used to put $\psi_\ell^2=0$. At the same time equation \eqref{eq2} implies that
$\psi_\ell^2$ is proportional to $(\dl{p}\cdot \dl{x}) \psi_\ell^1$ so that
$(\dl{p}\cdot \dl{x}) \psi_\ell^1=0$ and hence $\psi_\ell^1=0$ because its
component annihilated by $\dl{p}\cdot \dl{x}$ has been already gauged away.

\vspace{1cm}

%
%

\providecommand{\href}[2]{#2}\begingroup\raggedright\endgroup

\end{document}